\pdfoutput=1

\documentclass[11pt]{article}

\usepackage{graphicx,color,geometry}
\usepackage{amssymb,amsfonts,amsmath,amsthm,cite}

\newtheorem{theorem}   {Theorem}
\newtheorem{remark}   {Remark}

\begin{document}

\title{\bf Self-similar dynamics of morphogen gradients}

\author{Cyrill B. Muratov\thanks{Department of Mathematical Sciences,
    New Jersey Institute of Technology, Newark, NJ 07102, USA}, \and
  Peter V. Gordon\thanks{Department of Mathematical Sciences, New
    Jersey Institute of Technology, Newark, NJ 07102, USA} \and
  Stanislav Y. Shvartsman\thanks{Department of Chemical Engineering
    and Lewis Sigler Institute for Integrative Genomics, Princeton
    University, Princeton, NJ 08544, USA}}

\maketitle

\begin{abstract}
  We discovered a class of self-similar solutions in nonlinear models
  describing the formation of morphogen gradients, the concentration
  fields of molecules acting as spatial regulators of cell
  differention in developing tissues.  These models account for
  diffusion and self-induced degration of locally produced chemical
  signals.  When production starts, the signal concentration is equal
  to zero throughout the system. We found that in the limit of
  infinitely large signal production strength the solution of this
  problem is given by the product of the steady state concentration
  profile and a function of the diffusion similarity variable.  We
  derived a nonlinear boundary value problem satisfied by this
  function and used a variational approach to prove that this problem
  has a unique solution in a natural setting.  Using the asymptotic
  behavior of the solutions established by the analysis, we
  constructed these solutions numerically by the shooting method.
  Finally, we demonstrated that the obtained solutions may be easily
  approximated by simple analytical expressions, thus providing an
  accurate global characterization of the dynamics in an important
  class of non-linear models of morphogen gradient formation.  Our
  results illustrate the power of analytical approaches to studying
  nonlinear models of biophysical processes.
\end{abstract}

\section{Introduction}

Reaction-diffusion processes are involved in multiple aspects of
embryogenesis. In particular, a combination of extracellular diffusion
and degradation of locally produced proteins can establish
concentration fields of chemical signals that control spatial and
temporal gene expression patterns in developing tissues
\cite{ms02}. Such concentration fields are known as morphogen
gradients and have been identified in contexts as diverse as neural
development in vertebrates and wing morphogenesis in insects
\cite{tabata04,ab06}.

A canonical model of morphogen gradient formation is given by the
following initial boundary value problem
\cite{eldar03,twh07,othmer09,llnw09,wkg09}:
\begin{eqnarray}
 \label{eq:1}
 {\partial C \over \partial t} = D {\partial^2 C \over \partial x^2}
 - k(C) C, \qquad C(x, t = 0) = 0, \\
 \label{eq:2}
 -D \left. {\partial C \over \partial x} \right|_{x = 0} = Q, \qquad
 C(x = \infty, t) = 0.
\end{eqnarray}
Here $C = C(x, t)$ is the concentration of a morphogen as a function
of distance $x \geq 0$ to the tissue boundary and time $t \geq 0$. The
morphogen is produced with a constant rate $Q$ at the tissue boundary
($x = 0$), diffuses with diffusivity $D$ in the tissue ($x > 0$) and
is degraded in the tissue following some rate law characterized by the
pseudo first-order rate constant $k(C) > 0$. This model provides a
minimal description of complex biochemical and cellular processes in
real tissues, and has been recently used to quantitatively describe
morphogen gradients in a number of experimental systems
\cite{hswcjwhm10,kpbkbjg07,eldar03,ybnrpssb09}.

The level of gene expression in a particular cell located a certain
distance away from the signal source depends only on the local
concentration of that signal (or, more generally, on its time history
at that location). Some of the genes controlled by morphogen gradients
are directly involved in regulating processes of cell
differentiation. Another subset of genes act more indirectly,
regulating morphogens themselves. For example, morphogens can control
the expression of cell surface molecules that accelerate the rate of
morphogen degradation, establishing a feedback loop. Such self-induced
morphogen degradation can be modeled by having the degradation
constant $k(C)$ to be an increasing function of concentration.

The presence of cooperative effects that are commonly observed in the
transcriptional responses to morphogen gradients \cite{ab06} suggests
to consider a general class of degradation rates given by a power law
\cite{eldar03}:
\begin{eqnarray}
  \label{eq:kn}
  k(C) = k_n C^{n-1}, \qquad n > 1.
\end{eqnarray}
This type of nonlinearity was first considered by Eldar {\em et al.},
who demonstrated that power law degradation kinetics may generate
morphogen gradients that are robust with respect to large variations
in the source strength \cite{eldar03}. They based their conclusions on
the analysis of the steady version of Eq. \eqref{eq:1}. Specifically,
they demonstrated that, unlike the solutions of the corresponding
linear problem, i.e., Eqs. \eqref{eq:1} and \eqref{eq:2} with $k(C)
\equiv \text{const}$ (in which the solution depends on $Q$
multiplicatively), the stationary solution $C_s(x)$ of
Eqs. \eqref{eq:1}--\eqref{eq:kn} approaches an asymptotic limit when
$Q \to \infty$. As a consequence, the steady state of a system
operating in the regime of large $Q$'s will be insensitive to
variations in the strength of the source. This has important
implications for robustness of steady morphogen gradients established
by localized production, diffusion and self-induced degradation.

In this paper we show that robustness of the steady state solutions of
Eqs. \eqref{eq:1}--\eqref{eq:kn} discussed above carries over to the
solutions of the full time-dependent problem. Remarkably, we found
that for large values of $Q$ the solution of the initial boundary
value problem given by Eqs. \eqref{eq:1}--\eqref{eq:kn} approaches a
{\em self-similar} form:
\begin{eqnarray}
  \label{eq:Ct}
  C(x, t) = C_s(x) \phi(x / \sqrt{D t}),
\end{eqnarray}
where $\phi(\xi)$ is a universal function of $\xi = x / \sqrt{D t}$
which depends only on $n$ and decreases monotonically from $\phi = 1$
at $\xi = 0$ to $\phi = 0$ at $\xi = \infty$. The self-similar profile
function $\phi(\xi)$ is obtained by considering the singular version
of the initial boundary value problem with $Q = \infty$.

\begin{figure}
  \centering
    \includegraphics[width=3.25in]{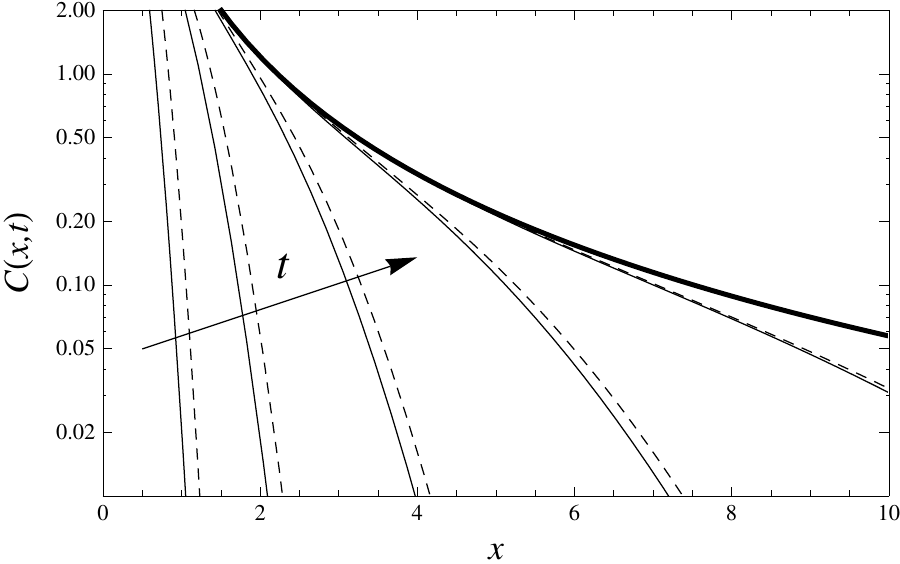}
    \caption{Comparison of the numerical solution of
      Eqs. \eqref{eq:1}--\eqref{eq:kn} for $n = 2$, $D = 1$, $k_2 = 1$
      and $Q = 10^3$ (thin solid lines) with the prediction of
      Eqs. \eqref{eq:C2} and \eqref{eq:w} (dashed lines) at times $t =
      0.04, 0.16, 0.64, 2.56, 10.24$. Thick solid line shows the
      steady state.  \label{f:C2}}
\end{figure}

Our results can be summarized as follows. We derived a nonlinear
boundary value problem satisfied by $\phi(\xi)$. We used a variational
approach to prove that this problem has a unique solution in a natural
setting. Using the asymptotic behavior of the solutions established by
the analysis, we constructed these solutions numerically by the
shooting method. We also showed that the obtained solutions may be
easily fitted to simple analytical expressions, thus providing an
accurate global characterization of the dynamics in an important class
of non-linear models of morphogen gradient formation. For example, in
the biophysically important case $n = 2$, in which the feedback is
mediated by the simplest bimolecular interaction our approach yields
\begin{eqnarray}
  \label{eq:C2}
  C(x, t) & \approx & {6  \phi(x / \sqrt{D t}) \over \left( x
      \sqrt{k_2 / D} + (12 \sqrt{k_2 D} / Q )^{1/3} \right)^2}, \\ 
  \label{eq:w}
  \phi(\xi) & \approx & {4000 + \xi^9 \over 4000 + 5 \xi^6 e^{\frac14 
      \xi^2}}, \qquad n = 2.
\end{eqnarray}
A comparison of the prediction of Eqs. \eqref{eq:C2} and \eqref{eq:w}
with the numerical solution of Eqs. \eqref{eq:1}--\eqref{eq:kn} for a
representative set of parameters is presented in Fig. \ref{f:C2}.

This paper is organized as follows. We first introduce the
non-dimensionalization of the considered initial boundary value
problem and give scaling arguments that lead us to consider singular
self-similar solutions of the problem. We then undertake a rigorous
analysis of existence, uniqueness and qualitative properties of the
considered self-similar solutions, which are summarized in Theorem
1. After that, we perform a numerical study of self-similar solutions
whose existence was established in the Analysis section and discuss
their implications to the dynamics. Finally, we discuss the obtained
results in connection with the studies of other types of self-similar
solutions in the considered problem, their significance for the
threshold crossing and conclude with some open problems.

\section{Results}
\label{sec:results}

In this section, we formulate a general setting for the analysis of
Eqs. \eqref{eq:1}--\eqref{eq:kn} and present our main findings.

\subsection{Non-dimensionalization}
\label{sec:non-dimens}

We begin by introducing the following dimensionless variables
\begin{eqnarray}
 \label{eq:42}
 x' = {x \over L}, \qquad t' = {t \over T}, \qquad u = {C \over C_0}, 
\end{eqnarray}
where
\begin{eqnarray}
 \label{eq:13}
 L = \left( {D \over k_n C_0^{n-1}} \right)^{1/2}, \quad T
 = {1 \over k_n C_0^{n-1}},
\end{eqnarray}
and $C_0$ is some reference morphogen concentration, corresponding,
e.g., to the threshold of expression of a downstream regulated
gene. In these new variables, the initial boundary value problem in
Eqs. \eqref{eq:1}--\eqref{eq:kn} takes the form
\begin{eqnarray}
  \label{eq:u}
  \left\{\begin{array}{ll}
      u_t=u_{xx}-u^n  &  (x,t)\in[0,\infty)\times(0,\infty), \\
      u_x(0, t)=-\alpha &  t\in(0,\infty),\\
      u(x, 0)=0 & x\in [0,\infty).
\end{array}\right.
\end{eqnarray}
where
\begin{eqnarray}
 \label{eq:16}
 \alpha = {Q \over \sqrt{D k_n C_0^{n+1}}}
\end{eqnarray}
is the dimensionless source strength.

\subsection{Scaling arguments}
\label{sec:scaling-arguments}

\begin{figure}[t]
  \centering
  \includegraphics[width=3in]{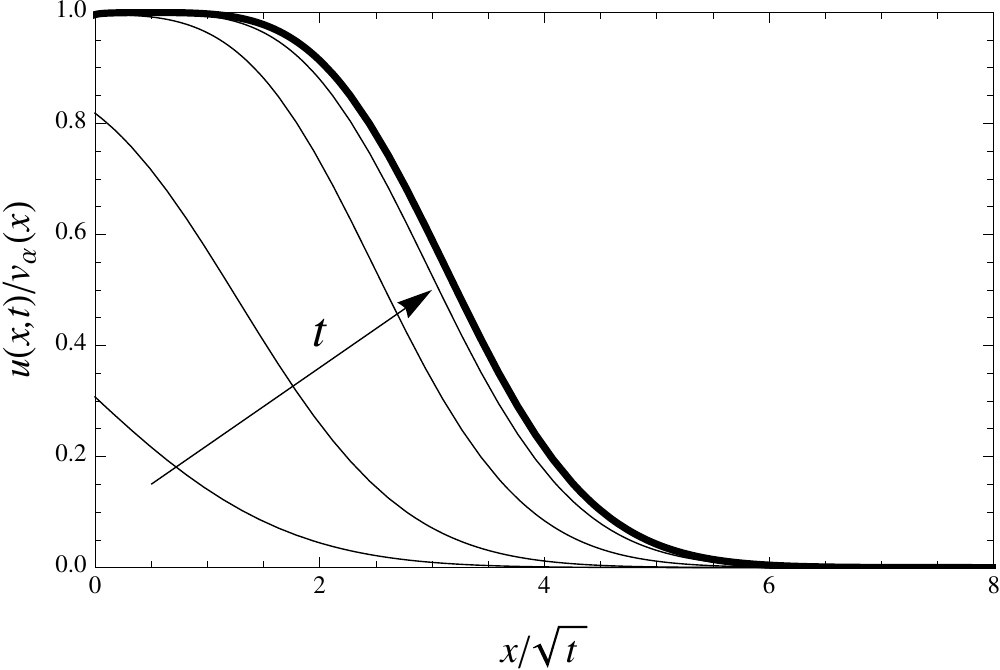}
  \caption{Example of the collapse of the solutions of
    Eq. \eqref{eq:u} onto a universal master curve at large
    times. Results of the numerical solution of Eq. \eqref{eq:u} with
    $n = 2$ and $\alpha = 1$. Thin lines show snapshots of the
    solution corresponding to $t = 0.1, 1, 10, 100$ (the direction of
    time increase is indicated by the arrow). The bold line shows the
    asymptotic master curve. \label{fig:collapse}}
\end{figure}

Let us now discuss the approach of the solutions of Eq. \eqref{eq:u}
to the unique steady state, which for this problem is given explicitly
by the following expression \cite{gsbms:pnas11}:
\begin{eqnarray}\label{eq:v}
  v_\alpha(x)=\left\{ \frac{2 (n+1)}{\bigl[ \left(2^n (n+1) \alpha
        ^{1-n}\right)^{\frac{1}{n+1}}+(n-1) x\bigr]^2} \right\}^{1
    \over n - 1}.    
\end{eqnarray}
It is not difficult to see that $u(x, t) \to v_\alpha(x)$ from below
as $t \to \infty$, implying that the fraction of the steady
concentration $u(x, t) / v_\alpha(x)$ reached at a given point $x \geq
0$ at time $t > 0$ will approach unity for $t \gg 1$
\cite{gsbms:pnas11}. In view of the diffusive nature of the processes
involved in establishing the steady concentration profile, one may
expect that the approach to the steady state occurs on the scale
associated with diffusion. Therefore, to better understand the
dynamics, we plot this fraction versus $x / \sqrt{t}$ for the solution
of Eq. \eqref{eq:u} with $n = 2$ and $\alpha = 1$ obtained numerically
for several values of $t$. The result is presented in
Fig. \ref{fig:collapse}. One can see from Fig. \ref{fig:collapse} that
the solution of Eq. \eqref{eq:u} at different values of $t$ collapses
onto a single master curve for $t \gg 1$. Furthermore, increasing the
value of $\alpha$ makes this collapse sooner. We also checked that the
same phenomenon occurs for different values of $n$. This strongly
suggests \cite{barenblatt} the existence of a hidden self-similarity
in the underlying dynamical behavior of the solutions of
Eq. \eqref{eq:u}.

We note that the solutions of Eq. \eqref{eq:u} are invariant with
respect to the following scaling transformation:
\begin{eqnarray}
  \label{eq:xtu}
  \alpha' = \lambda \alpha, \quad t' = \lambda^{2(1-n) \over 1+n} t, \quad x' 
  = \lambda^{1-n \over n+1} x, \quad u' = \lambda^{2 \over n+1} u. 
\end{eqnarray}
In other words, increasing the source strength $\alpha$ by a factor of
$\lambda$ decreases the time scale of approach to the steady state by
a factor of $\lambda^{2 (n - 1) \over n + 1}$ at fixed value of $x /
\sqrt{t}$. Therefore, the approach to the universal curve in
Fig. \ref{fig:collapse} must occur on the time scale
\begin{eqnarray}
  \label{eq:5}
  \tau_n \sim \alpha^{2(1-n) \over n+1}.   
\end{eqnarray}
This scale was recently identified by us in the analysis of the local
accumulation time in the particular case of Eq. \eqref{eq:u}
\cite{gsbms:pnas11}. Observe that $\tau_n \to 0$ as $\alpha \to
\infty$ for all $n > 1$. Thus, our numerical results suggest that in
the limit $\alpha \to \infty$ the ratio $u(x, t) / v_\alpha(x)$
depends only on $x / \sqrt{t}$ for all $t > 0$, exhibiting
self-similar behavior.

\subsection{Singular solutions and the similarity ansatz}
\label{sec:similarity-ansatz}

The numerical observations discussed in the preceding section suggest
the need to consider the following singular initial boundary value
problem:
\begin{eqnarray}
  \label{eq:uu}
  \left\{\begin{array}{ll}
      u_t=u_{xx}-u^n  &  (x, t)\in(0,\infty)\times(0,\infty), \\
      u(0,t)=\infty,  &  t\in(0,\infty),\\
      u(x,0)=0 & x\in (0,\infty).
\end{array}\right.
\end{eqnarray}
By a solution to Eq. \eqref{eq:uu}, we mean a classical solution for
all $(x, t) \in (0, \infty) \times (0, \infty)$ decaying sufficiently
fast as $x \to +\infty$ for all $t > 0$, and continuous up to $t = 0$
for all $x > 0$. Note that for each $n > 1$ this problem possesses a
{\em singular} stationary solution
\begin{eqnarray}
  \label{eq:vinf}
  v_\infty(x) = \left( \frac{2 (n+1) }{(n-1)^2} \right)^{1
    \over n - 1} \left( {1 \over x} \right)^{2 \over n - 1},
\end{eqnarray}
which is the limit of $v_\alpha(x)$ as $\alpha \to \infty$ for each $x
> 0$. Therefore, in view of the discussion above, solution of
Eq. \eqref{eq:uu} is expected to take the form
\begin{eqnarray}
  \label{eq:simanz}
  u(x, t) = v_\infty(x) \phi(x / \sqrt{t}),
\end{eqnarray}
for some universal function $\phi(\xi)$ with values between zero and
one, which depends only on $n$.

We now substitute the similarity ansatz from Eq. \eqref{eq:simanz}
into Eq. \eqref{eq:uu}. After some algebra, this leads to the
following equation for the self-similar profile $\phi$:
\begin{eqnarray}
  \label{eq:phixi}
  \xi^2 {d^2 \phi \over d \xi^2} + \left( {\xi^3 \over 2} - {4 \xi
      \over n - 1} \right) {d \phi \over d \xi} + {2 (n + 1) \over (n
    - 1)^2} \phi (1 - \phi^{n-1}) = 0,
\end{eqnarray}
which must hold for all $\xi \in (0, \infty)$. Consistent with the
interpretation of Eq. \eqref{eq:uu}, this equation needs to be
supplemented with the boundary-like conditions
\begin{eqnarray}
  \label{eq:phibc}
  \lim_{\xi \to 0} \phi(\xi) = 1, \qquad \lim_{\xi \to \infty}
  \phi(\xi) = 0. 
\end{eqnarray}

Existence and multiplicity of solutions of Eqs. \eqref{eq:phixi},
\eqref{eq:phibc} are not {\em a priori} obvious in view of both the
non-linearity and the presence of singular terms in the considered
boundary value problem. Below we establish existence and uniqueness of
these solutions for all $n > 1$ within a natural class of
functions. Later, in the following section, we construct these
solutions numerically.

\subsection{Analysis}
\label{sec:analysis}

This section is concerned with the proof of existence and uniqueness
of solutions of Eqs. \eqref{eq:phixi}, \eqref{eq:phibc} in a natural
setting. The reader interested in the application of our analysis to
Eqs. \eqref{eq:1}--\eqref{eq:kn} may skip this part and proceed
directly to the following section, which discusses the numerical
solution of Eqs. \eqref{eq:phixi}, \eqref{eq:phibc}.

For the purposes of the analysis it is convenient to rewrite
Eq. \eqref{eq:phixi}, using a new variable $\zeta = \ln \xi$:
\begin{eqnarray} 
  \label{eq:phiz} {d^2 \phi \over d \zeta^2} +
  \left(\frac{e^{2\zeta}}{2} - \frac{n+3}{n-1}\right){d \phi \over d
    \zeta} + 
  \frac{2(n+1)}{(n-1)^2}\phi(1-\phi^{n-1})=0,
\end{eqnarray}
where $\phi \in C^2(\mathbb R)$ and has the respective limits
\begin{eqnarray}
  \label{eq:phibcz}
  \lim_{\zeta \to -\infty} \phi(\zeta) = 1, \qquad \lim_{\zeta \to +\infty}
  \phi(\zeta) = 0. 
\end{eqnarray}

We will prove existence and uniqueness of solutions of
Eqs. \eqref{eq:phiz} and \eqref{eq:phibcz} in the weighted Sobolev
space $H^1(\mathbb R, d \mu)$, which is obtained as the completion
of the family of smooth functions with compact support with respect to
the Sobolev norm $||.||_{H^1(\mathbb R, d \mu)}$, defined as
\begin{eqnarray}
  \label{eq:H1}
  ||w||_{H^1(\mathbb R, d \mu)}^2  =  ||w_\zeta||_{L^2(\mathbb R, d
    \mu)}^2 +  ||w||_{L^2(\mathbb R, d \mu)}^2,
\end{eqnarray}
where $||w||_{L^2(\mathbb R, d \mu)}^2 = \int_\mathbb{R} w^2(\zeta)
d \mu(\zeta)$, and the measure $d \mu$ is
\begin{eqnarray}
  \label{eq:muu}
  d \mu(\zeta) = \rho(\zeta) d \zeta, \qquad \rho(\zeta) =  
  \exp\left\{\frac{e^{2\zeta}}{4}-\left(\frac{n+3}{n-1}\right) \zeta 
  \right\}.
\end{eqnarray}
Our existence and uniqueness result is given by the following
theorem. 

\begin{theorem}
  There exists a unique weak solution $\phi$ of Eq. \eqref{eq:phiz},
  such that $\phi - \eta \in H^1(\mathbb R, d \mu)$, with $\mu$
  defined in Eq. \eqref{eq:muu}, for every $\eta \in C^\infty(\mathbb
  R)$, such that $\eta(\zeta) = 1$ for all $\zeta \leq 0$ and
  $\eta(\zeta) = 0$ for all $\zeta \geq 1$. Furthermore, $\phi \in
  C^\infty(\mathbb R)$, satisfies Eq. \eqref{eq:phiz} classically and
  $0 < \phi < 1$. In addition, $\phi$ is strictly decreasing and
  satisfies Eq. \eqref{eq:phibcz}.
\end{theorem}

{\em \bf \small \noindent Proof. } The proof consists of five
steps. \smallskip

  {\bf \small \noindent Step 1.} We first note that
  Eq. \eqref{eq:phiz} is the Euler-Lagrange equation for the energy
  functional 
  \begin{align} 
    \label{eq:E}
    {\cal E}(\phi)=\int_\mathbb{R} \Biggl\{ \frac{1}{2} \left({d \phi
        \over d \zeta} \right)^2 + {\eta \over n - 1}
    -\frac{\phi^2(n+1-2\phi^{n-1})}{(n-1)^2} \Biggr\} d \mu,
\end{align}
where $\eta(\zeta)$ is as in the statement of the theorem. The natural
admissible class $\mathcal A$ for $\mathcal E$ is:
\begin{eqnarray}
  \label{eq:A}
  \mathcal A := \{ \phi \in H^1_\mathrm{loc}(\mathbb R): \phi - \eta
  \in H^1(\mathbb R, d \mu), \hspace{1mm} 0 \leq \phi \leq 1 \}.  
\end{eqnarray}
Note that the role of $\eta$ in the definition of $\mathcal E$ is to
ensure that the integral in Eq. \eqref{eq:E} converges for all $\phi
\in \mathcal A$. The precise form of $\eta(\zeta)$ is unimportant.

\smallskip

{\bf \small \noindent Step 2.} We now establish weak sequential
lower-semicontinuity and coercivity of the functional $\mathcal E$ in
the admissible class $\mathcal A$ in the following sense: let $\phi_k
= \eta + w_k$, where $w_k \rightharpoonup w$ in $H^1(\mathbb R, d
\mu)$. Then 1) $\liminf_{k \to \infty} \mathcal E(\phi_k) \geq
\mathcal E(\phi)$, where $\phi = \eta + w$, and 2) if $\mathcal
E(\phi_k) \leq M$ for some $M \in \mathbb R$, then
$||w_k||_{H^1(\mathbb R, d \mu)} \leq M'$ for some $M' > 0$.

Let us introduce the notation $\mathcal E(\phi, (a, b))$ for the
integral in Eq. \eqref{eq:E}, in which integration is over all $\zeta
\in (a, b)$. Arguing as in \cite[Lemma 4.1]{lmn:cpam04}, for $R \gg 1$
we have
\begin{eqnarray}
  \label{eq:poinc}
  \int_R^\infty w^2 d \mu \leq 2 e^{-2 R} \int_R^\infty
  w_\zeta^2 d \mu \qquad \forall w \in H^1(\mathbb R, d
  \mu).
\end{eqnarray}
Therefore, from Eq. \eqref{eq:poinc} we find that
\begin{eqnarray}
  \label{eq:ERp}
  \mathcal E(\phi_k, (R, +\infty)) \geq \left( e^{2 R} - {n + 1 \over
      (n - 1)^2} \right) \int_R^\infty w_k^2 d \mu,
\end{eqnarray}
which is positive for $R \gg 1$. Similarly, taking into account that
the integrand in Eq. \eqref{eq:E} is non-negative for $\zeta \leq 0$,
we have $\mathcal E(\phi_k, (-\infty, 0)) \geq 0$. Since $\mathcal
E(\cdot, (0, R))$ is lower-semicontinuous by standard theory
\cite{dalmaso}, we obtain $\mathcal E(\phi_k) \geq \mathcal E(\phi_k,
(0, R))$, yielding the first claim in view of arbitrariness of $R \gg
1$.

To prove coercivity, we first note that by Eq. \eqref{eq:poinc}
\begin{eqnarray}
  \label{eq:Ec}
  \mathcal E(\phi_k, (R, +\infty)) \geq \int_R^\infty \left\{ \frac12
    \left( { d w_k \over d \zeta} \right)^2 - {n + 1 \over (n
      - 1)^2} w_k^2 \right\}  d \mu \nonumber \\
  \geq \frac14 \int_R^\infty \left\{ \left( { d w_k \over d \zeta}
    \right)^2 + w_k^2 \right\} d \mu,
\end{eqnarray}
for $R \gg 1$. On the other hand, since $n - 1 - \phi^2 (n + 1 - 2
\phi^{n-1}) \geq (n - 1) (1 - \phi)^2$ for all $0 \leq \phi \leq 1$,
we have
\begin{eqnarray}
  \label{eq:Emc}
  \mathcal E(\phi_k, (-\infty, 0)) \geq
  \int_{-\infty}^0 \left\{ \frac12 \left( { d w_k \over d \zeta}
    \right)^2 + {w_k^2 \over n - 1} \right\} d \mu. 
\end{eqnarray}
Finally, by boundedness of $\phi_k$ and $\eta$, we also have
\begin{eqnarray}
  \label{eq:E0c}
  \mathcal
  E(\phi_k, (0, R)) \geq \frac14 \int_0^R \left\{ \left(
      { d w_k \over d \zeta} \right)^2 + w_k^2 \right\} d
  \mu - C R, 
\end{eqnarray}
for some $C > 0$, whenever $||w_k||_{H^1(\mathbb R, d \mu)}$ is
sufficiently large. So the second claim follows.

\smallskip

{\bf \small \noindent Step 3.} In view of the lower-semicontinuity and
coercivity of $\mathcal E$ proved in Step 2, by the direct method of
calculus of variations there exists a minimizer $\phi \in \mathcal A$
of $\mathcal E$. Noting that $\phi = 0$ and $\phi = 1$ solve
Eq. \eqref{eq:phiz}, we also have that $\phi$ is a weak solution of
Eq. \eqref{eq:phiz} by continuous differentiability of $\mathcal E$ in
$H^1(\mathbb R, d \mu)$. Furthermore, by standard elliptic regularity
theory \cite{gilbarg}, $\phi \in C^\infty(\mathbb R)$ and is, in fact,
a classical solution of Eq. \eqref{eq:phiz}. Also, by strong maximum
principle \cite{gilbarg}, we have $0 < \phi < 1$. To show
monotonicity, suppose, to the contrary, that $\phi(a) < \phi(b)$ for
some $a < b$. Then $\phi(\zeta)$ attains a local minimum for some
$\zeta_0 \in (-\infty, b)$. However, by Eq. \eqref{eq:phiz} we have
$d^2 \phi(\zeta_0) / d \zeta^2 < 0$, giving a contradiction. By the
same argument $d \phi / d \zeta = 0$ is also impossible for any $\zeta
\in \mathbb R$. Finally, since $\phi - \eta \in H^1(\mathbb R, d
\mu)$, monotonicity implies Eq. \eqref{eq:phibcz}.

\smallskip

{\bf \small \noindent Step 4.} We now discuss the asymptotic behavior
of the minimizers obtained in Step 3 as $\zeta \to
+\infty$. Performing the Liouville transformation by introducing $\psi
= \phi \sqrt{\rho} \in L^2(R, +\infty)$, where $\rho$ is defined in
Eq. \eqref{eq:muu} and $R \in \mathbb R$ is arbitrary, we rewrite
Eq. \eqref{eq:phiz} in the form
\begin{eqnarray}
  \label{eq:schr}
  {d^2 \psi \over d \zeta^2} = q(\zeta) \psi, \qquad \zeta \geq R.  
\end{eqnarray}
Here $q(\zeta) = q_0(\zeta) +
q_1(\zeta)$, where
\begin{align}
  \label{eq:q0}
  q_0(\zeta) & = \frac14\left(
    \frac{e^{4\zeta}}{4}+\frac{n-5}{n-1}e^{2\zeta} +1
  \right), \\ 
  \label{eq:q1}
  q_1(\zeta) & = \frac{2(n+1)}{(n-1)^2} \, \phi^{n-1} (\zeta). 
\end{align}
Observe that $q(\zeta) \geq q_0(\zeta) \geq \tfrac14 > 0$ for all
$\zeta \geq R \gg 1$. Therefore, Eq. \eqref{eq:schr} has two
linearly-independent positive solutions $\psi_1$ and $\psi_2$, such
that $\psi_1 \to 0$ and $\psi_2 \to \infty$ together with their
derivatives as $\zeta \to +\infty$ (see e.g. \cite{sansone}). In
particular, $\psi = C \psi_1 \in L^2(R, +\infty)$ for some $C > 0$, and
\begin{eqnarray}
  \label{eq:q11}
  q_1(\zeta) = o( \rho^{1 - n \over 2}),
\end{eqnarray}
so $q_1(\zeta)$ has a super-exponential decay as $\zeta \to +\infty$.

Now let $\psi_0$ be the unique positive solution of
Eq. \eqref{eq:schr} with $q = q_0$ and $\psi_0(R) = 1$ which goes to
zero as $\zeta \to +\infty$. Then we claim that $\psi_1(\zeta) /
\psi_0(\zeta) \to c$ for some $0 < c < \infty$. Indeed, after
straightforward algebraic manipulations we have
\begin{eqnarray}
  \label{eq:log}
  {d \over d \zeta} \ln \left( {\psi_1 \over \psi_0} \right) = -
  \int_\zeta^{\infty} q_1(s) {\psi_1(s) \psi_0(s) \over \psi_1(\zeta)
    \psi_0(\zeta)} ds.  
\end{eqnarray}
The result then follows by the decay of $\psi_0$ and $\psi_1$ and the
estimate in Eq. \eqref{eq:q11} upon integration of Eq. \eqref{eq:log}.

\smallskip

{\bf \small \noindent Step 5.} We now prove uniqueness of the obtained
solution, taking advantage of a sort of convexity of $\mathcal E$
pointed out in \cite{kawohl85}. Suppose, to the contrary, that there
are two functions $\phi_1, \phi_2 \in \mathcal A$ which solve
Eq. \eqref{eq:phiz} weakly. Define $\phi^t = \sqrt{t \phi_2^2 + (1 -
  t) \phi_1^2} \in \mathcal A$, since in view of the result of Step 4
we have $m < \phi_1 / \phi_2 < M$ for some $M > m > 0$.  It is easy to
see that the function $E(t) := \mathcal E(\phi^t)$ is twice
continuously differentiable for all $t \in [0, 1]$. A direct
computation yields
\begin{eqnarray}
  \label{eq:dE2}
  {d^2 E(t) \over d t^2} = \int_\mathbb{R} \Biggl\{ {\phi_1^2 \phi_2^2
    \over (t \phi_2^2 + (1 - t) \phi_1^2)^3 } \left( \phi_2 {d \phi_1
      \over d \zeta} - \phi_1 {d \phi_2 \over d \zeta}\right)^2
  \nonumber \\
  + {n + 1 \over 2 n - 2} (\phi_1^2 - \phi_2^2)^2  (t \phi_2^2 + (1
  - t) \phi_1^2)^{n - 3 \over 2} \Biggr\} d \mu(\zeta). 
\end{eqnarray}
Therefore, $d^2 E(t) / dt^2 > 0$ for all $t \in [0,1]$, and so $E(t)$
is strictly convex. However, since the map $t \mapsto \phi^t - \eta$
is of class $C^1([0, 1]; H^1(\mathbb R, d \mu))$, this
contradicts the fact that $dE(0)/dt = dE(1)/dt = 0$ by the assumption
that $\phi_1$ and $\phi_2$ solve weakly Eq. \eqref{eq:phiz} and hence
are critical points of $\mathcal E$. ~$\Box$

\begin{remark}
  \label{r:1}
  Note that the arguments in Step 4 above imply that the asymptotic
  behavior of the self-similar profile $\phi(\xi)$ as $\xi \to \infty$
  is the same as that of the decaying solution of Eq. \eqref{eq:phixi}
  linearized around $\phi = 0$. A similar argument shows that the
  behavior of $\phi(\xi)$ as $\xi \to 0$ is the same as that of the
  corresponding solution of Eq. \eqref{eq:phixi} linearized around
  $\phi = 1$.
\end{remark}

\subsection{Numerics}
\label{sec:numerics}

We now construct the self-similar profiles, whose existence and
uniqueness was established in Theorem 1, numerically for several
values of $n > 1$. We use the shooting method to construct the
solutions of Eq. \eqref{eq:phixi}, which requires knowledge of the
asymptotic behavior of $\phi(\xi)$ near $\xi = 0$ and $\xi =
\infty$. To obtain this behavior, we linearize Eq. \eqref{eq:phixi}
around the equilibria $\phi = 0$ and $\phi = 1$, which is justified by
Remark \ref{r:1}. Denote the corresponding solutions of the linearized
equations as $\phi_0$ and $\phi_1$, respectively.  By a direct
computation
\begin{eqnarray}
  \label{eq:phi0}
  \phi_0(\xi) = C_1 \xi^{2 \over n-1} M\left(
    {1 \over n - 1}, {1 \over 2}, -{\xi^2 \over 4} \right) \nonumber
  \qquad \qquad \qquad \\ 
  + C_2 e^{-\frac14 \xi^2} \xi^{2 \over n - 1} U
  \left(\frac12 +  {1 \over 1 - n}, {1 \over 2}, {\xi^2 \over 4}
  \right), 
\end{eqnarray}
where $M(a, b, z)$ and $U(a, b, z)$ are the confluent hypergeometric
functions of the first and second kind, respectively
\cite{abramowitz}. Using the asymptotic expansions of these functions
for large $z$ \cite{abramowitz}, one can see that $\phi_0(\xi) \to 0$
as $\xi \to \infty$, if and only if the constant $C_1 = 0$. Therefore,
from the asymptotic expansion of $U$ we have
\begin{eqnarray}
  \label{eq:phi00}
  \phi(\xi) \sim e^{-\frac14 \xi^2} \xi^{5 - n \over n - 1}, \qquad
  \xi \to \infty.
\end{eqnarray}
Similarly
\begin{eqnarray}
  \label{eq:phi1}
  \phi_1(\xi) = \xi^{2(n+1) \over n - 1} \Biggl\{ C_1 M \left({n + 1
      \over n - 1}, {5 n - 1 \over 2 n - 2}, -{\xi^2 \over 4} \right)
  \nonumber \\ 
  + C_2 U \left({n + 1 
      \over n - 1} , {5 n - 1 \over 2 n - 2}, -{\xi^2 \over 4} \right)
  \Biggr\}.  
\end{eqnarray}
Once again, for a bounded solution at $\xi = 0$ we must set $C_2 = 0$,
which leads to
\begin{eqnarray}
  \label{eq:6}
  1 - \phi(\xi) \sim \xi^{2(n + 1) \over n - 1}, \qquad \xi \to 0.
\end{eqnarray}

\begin{figure}[t]
  \centering
  \includegraphics[width=3in]{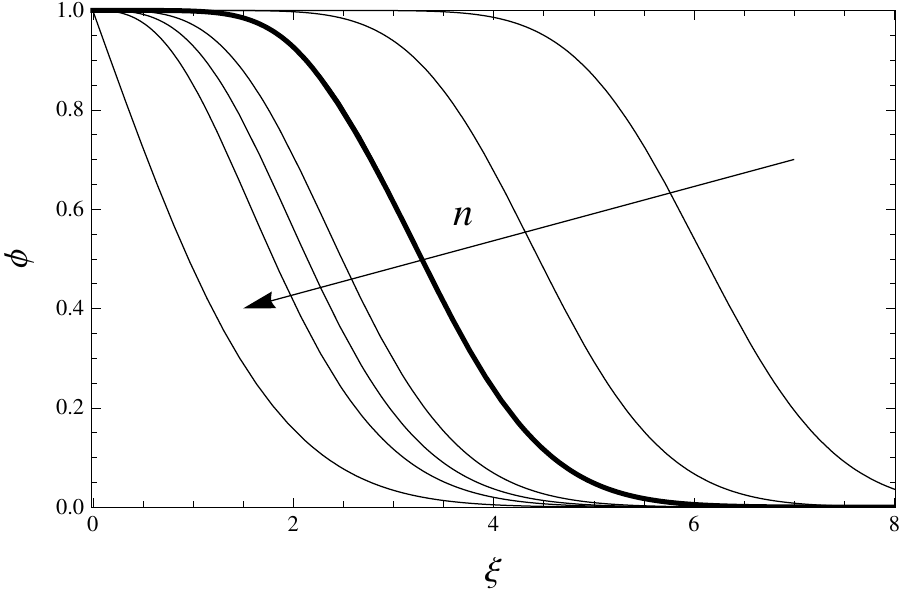}
  \caption{Self-similar profiles $\phi(\xi)$ for different values of
    $n$. Results of the numerical solution of Eqs. \eqref{eq:phixi}
    and \eqref{eq:phibc} for $n = 1.25, 1.5, 2, 3, 4, 6, \infty$. The
    thick line is the graph of the function given by Eq. \eqref{eq:w}
    overlaying the profile for $n = 2$.  \label{fig:selfsim_all}}
\end{figure}

The results of the numerical solution of Eq. \eqref{eq:phixi} whose
asymptotic behavior is governed by Eqs. \eqref{eq:phi0} and
\eqref{eq:phi1} are presented in Fig. \ref{fig:selfsim_all}. One can
see that the self-similar profiles form a monotonically decreasing
family of functions parametrized by $n$. The solutions $\phi(\xi)$
approach $\overline\phi(\xi) = 1$ on finite intervals as $n \to 1$ and
$\underline{\phi}(\xi) = 1 - \mathrm{erf}(\xi / 2)$ as $n \to \infty$
(the latter solves Eq. \eqref{eq:phixi} corresponding to $n =
\infty$). We also found that for the biophysically important case $n =
2$ the self-similar profile can be approximated by the simple
expression given by Eq. \eqref{eq:w} within $\sim 1\%$ accuracy. The
graph of this function, which essentially coincides with that of the
numerical solution of Eq. \eqref{eq:phixi} is shown in
Fig. \ref{fig:selfsim_all} with a thick line. Note that this profile
also coincides with the limiting profile in Fig. \ref{fig:collapse}
for $t = \infty$.

\begin{figure}
  \centering
  \includegraphics[width=3in]{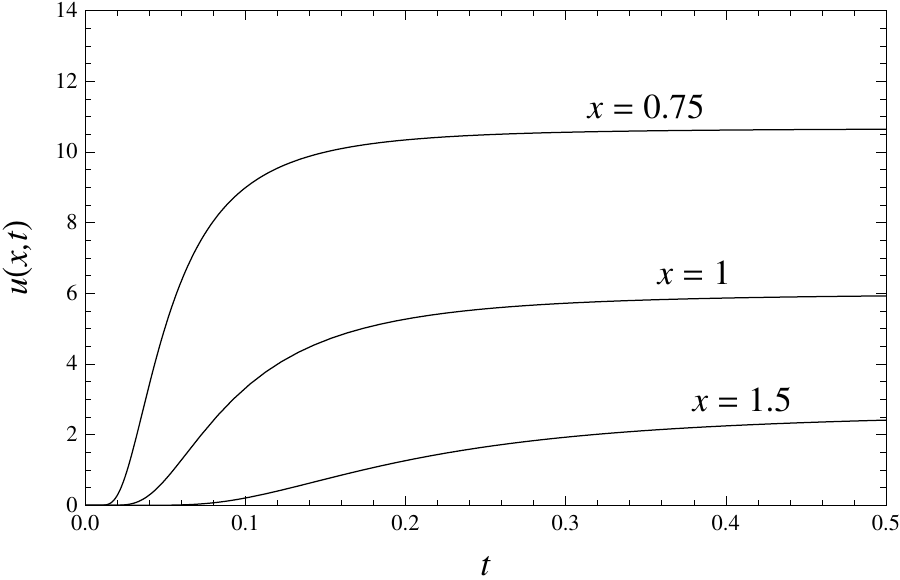}
  \caption{Self-similar solutions $u(x, t)$ of Eq. \eqref{eq:uu} at
    several values of $x$ for $n = 2$.  \label{fig:uxt}}
\end{figure}

\subsection{Dynamics}
\label{sec:dynam}

We now discuss the dynamical behavior of the obtained self-similar
solutions of Eq. \eqref{eq:uu}. The picture remains qualitatively the
same for all $n > 1$, so in the following we restrict our attention to
the biophysically important case of $n = 2$. 

First consider the time course of the solution $u(x, t)$ given by
Eq. \eqref{eq:simanz} at a fixed location, i.e. at a fixed value of $x
> 0$. From the self-similarity ansatz in Eq. \eqref{eq:simanz} it is
clear that the time scale of these dynamics is governed by diffusion,
i.e., $t \sim x^2$. A convenient characterization of local dynamical
time scale can be made in terms of the local accumulation time
$\tau^\infty(x) = \int_0^\infty t p(x, t) dt$, where the probability
density-like quantity $p(x, t) = {1 \over v_\infty(x)} {\partial u(x,
  t) \over \partial t}$ \cite{berezhkovskii10,gsbms:pnas11}. Upon
substitution of Eq. \eqref{eq:simanz} into this formula and an
integration by parts, one obtains
\begin{eqnarray}
  \label{eq:tau}
  \tau^\infty(x) = a x^2, \qquad a = 2 \int_0^\infty \xi^{-3} (1 -
  \phi(\xi)) d \xi,   
\end{eqnarray}
where numerically $a \simeq 0.122$. We note that by Eq. \eqref{eq:6}
the integral in Eq. \eqref{eq:tau} converges for all $n > 1$.  The
solution for several values of $x$ is shown in
Fig. \ref{fig:uxt}. Furthermore, as follows from Eqs. \eqref{eq:phi00}
and \eqref{eq:6}, when $t \ll \tau^\infty(x)$, we have $u(x, t) \sim
(x / t^{3/2} ) e^{-\frac{x^2}{4 t}}$, which is exponentially small. At
the same time, for $t \gg \tau^\infty(x)$ we have $(v_\infty(x) - u(x,
t)) / v_\infty(x) \sim (\tau^\infty(x) / t)^3$, i.e., $u$ approaches
the stationary solution, with the distance to the stationary solution
decaying as $O(t^{-3})$.

\begin{figure}
  \centering
  \includegraphics[width=3in]{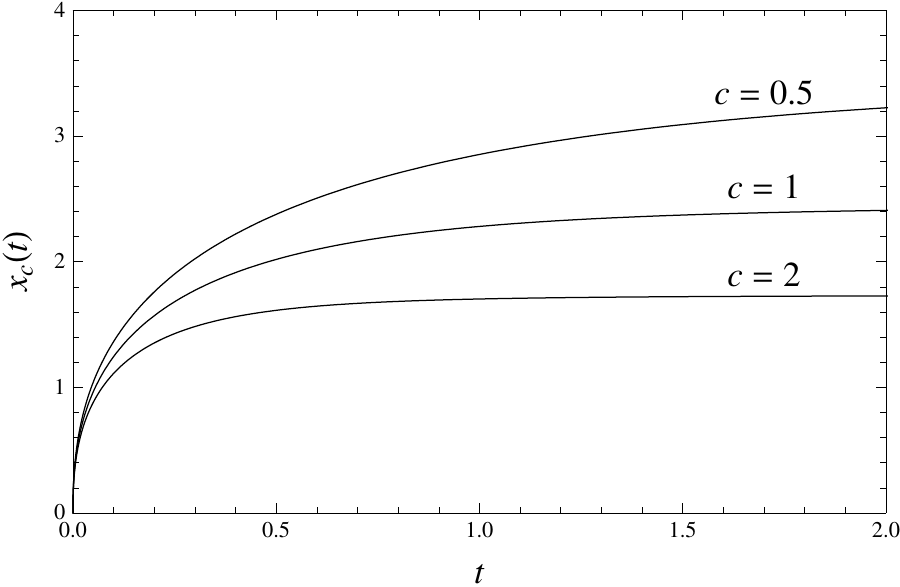}
  \caption{The positions $x_c(t)$ of level sets $\{u(x, t) = c\}$ of
    the self-similar solution of Eq. \eqref{eq:uu} at several values
    of $c$ for $n = 2$.  \label{fig:xu}}
\end{figure}

We now consider the motion of the level sets of the solutions of
Eq. \eqref{eq:uu}. For a given $c > 0$, let us define $x_c(t)$ as the
unique value of $x$, such that $u(x, t) = c$ for each $t > 0$. As
follows from Eqs. \eqref{eq:simanz}, the function $x_c(t)$ can be
determined parametrically as
\begin{eqnarray}
  \label{eq:xc}
  x_c = (6 \phi(\xi) / c )^{1/2}, \qquad t =
  6 \phi(\xi) / (c\xi^2), \qquad n = 2. 
\end{eqnarray}
The graphs of $x_c(t)$ for a few values of $c$ are shown in
Fig. \ref{fig:xu}. Once again, the dynamics of $x_c$ can be
characterized by the local accumulation time $\tau^\infty(x_c^\infty)$
given by Eq. \eqref{eq:tau}, where $x_c^\infty = (6 / c)^{1/2}$ is the
asymptotic value of $x_c(t)$ as $t \to \infty$. One can see from
Eqs. \eqref{eq:phi00} and \eqref{eq:xc} that for $t \ll
\tau^\infty(x_c^\infty)$ we have $x_c \simeq 2 (t \ln t^{-1}
)^{1/2}$. Thus, all level sets move together for short times, as can
also be seen from Fig. \ref{fig:xu}. On the other hand, for $t \gg
\tau^\infty(x_c^\infty)$ the level set position $x_c(t)$ approaches
$x_c^\infty$ as $x_c^\infty - x_c(t) = O(t^{-3})$. Within $\sim 2\%$
accuracy the functions $x_c(t)$ can be approximated by the following
simple expression:
\begin{eqnarray}
  \label{eq:xcfit}
  x_c(t) \approx \left( {4 t \ln [3.2 + 6 /(c t)] \over 1 + 0.76 c t}
  \right)^{1/2}, \qquad n = 2.
\end{eqnarray}
This formula implies that $x_c(t)$ comes within $5\%$ of $x_c^\infty$
at $t \simeq 2 \tau^\infty(x_c^\infty)$.

\section{Discussion}
\label{sec:discussion}

We have provided an analytical characterization of the dynamics of
morphogen concentration profiles in models with self-induced morphogen
degradation. Our results reveal the presence of self-similarity in the
course of the approach of the concentration profiles to their steady
states in either the limit of large source strengths or for large
distances away from the source. In addition to demonstrating the
self-similar nature of the dynamics, we rigorously established
existence and uniqueness of the associated singular self-similar
solutions to Eq. \eqref{eq:1} and constructed these solutions
numerically for several values of $n$. The obtained solutions may be
readily used to study various characteristics of the local kinetics of
morphogen concentration. In particular, Eqs. \eqref{eq:xc} and
\eqref{eq:xcfit} obtained from the numerical self-similar solutions
provide a characterization of threshold crossing events, which
determine the times at which a morphogen gradient switches the gene
expression on or off at a given point.

\subsection{Comparison with self-similar transients} 
\label{sec:comp-with-meas}

Mathematically, we have constructed a new class of self-similar
solutions to Eqs. \eqref{eq:1}, \eqref{eq:kn} on half-line. These
solutions can be trivially extended to the whole real line by a
reflection and can be viewed as singular solutions of
Eqs. \eqref{eq:1}, \eqref{eq:kn} that blow up at the origin. We note
that Eqs. \eqref{eq:1}, \eqref{eq:kn} and their $d$-dimensional analog
is known to possess another kind of self-similar solutions which were
extensively studied \cite{brezis83,kamin85,oswald88}, starting from
the works of \cite{galaktionov85,brezis86} (see also \cite{bernoff10}
for a less technical introduction and \cite{escobedo87} for a
variational treatment). These are classical solutions of
Eqs. \eqref{eq:1}, \eqref{eq:kn} for $t > 0$, which concentrate to a
point mass at $t = 0$ and describe the {\em transient} dynamics for
such initial data. They also play an important role in the long-time
behavior of the associated initial value problem (see e. g.
\cite{galaktionov85,escobedo95,bricmont96,wayne97,herraiz99}).  The
solutions found by us can be viewed as the counterparts of the {\em
  very singular solutions} constructed in \cite{brezis86}. Our
solutions are more singular than those of \cite{brezis86} in the sense
that the singularity in the former is concentrated on a half-line ($x
= 0, t > 0$) in the $(x, t)$ plane, while the singularity in the
latter occurs only at a single point ($x = 0, t = 0$).

\subsection{Extensions and open problems}
\label{sec:extensions}

It would be interesting to understand the role our self-similar
solutions play for the singular solutions of the initial value problem
associated withEqs. \eqref{eq:1}, \eqref{eq:kn} for general non-zero
initial data. Let us point out that even the basic questions of
existence and uniqueness of such singular solutions for the considered
parabolic problems in suitable function classes are currently open
(see \cite{veron11} for a very recent related work). Other natural
extensions include higher dimensional versions of the considered
problem, as well as a proof of global stability of self-similar
solutions. From the point of view of applications, it is also
important to consider singular solutions of Eq. \eqref{eq:1} for
time-varying sources, for which both types of self-similar solutions
that are present in the system may be relevant.

\paragraph{Acknowledgements.} The work of PVG was supported, in part,
by the United States -- Israel Binational Science Foundation grant
2006-151. CBM acknowledges partial support by NSF via grants
DMS-0718027 and DMS-0908279. SYS acknowledges partial support by NSF
via grant DMS-0718604 and by NIH via grant GM078079. CBM and PVG would
like to acknowledge valuable discussions with V. Moroz.

\bibliographystyle{unsrt}

\bibliography{../mura,../nonlin,../bio,../egfr}

\end{document}